\documentclass[aps,pre,twocolumn,showpacs,groupedaddress,superscriptaddress]{revtex4-1}
\usepackage{graphicx} 
\usepackage{amsmath}
\usepackage{color}
\usepackage{epsfig}
\usepackage{epstopdf}

\begin{document} %{{{ 

\title{Mod-CSA: Modularity optimization by conformational space annealing}
\author{Juyong Lee}
\email{juyong@kias.re.kr}
\affiliation{School of Computational Sciences, Korea Institute of Advanced Study, Seoul, Korea}
\author{Steven P. Gross}
\email{sgross@uci.edu}
\affiliation{Department of Developmental and Cell Biology, University of California, Irvine, USA}
\author{Jooyoung Lee}
\email{jlee@kias.re.kr}
\affiliation{School of Computational Sciences, Korea Institute of Advanced Study, Seoul, Korea}
\date{\today} %}}} 

\begin{abstract} %{{{
We propose a new modularity optimization method, Mod-CSA, based on stochastic global optimization algorithm, conformational space annealing (CSA).
Our method outperforms simulated annealing in terms of both efficiency and accuracy, finding higher modularity partitions with less computational resources required. 
The high modularity values found by our method are higher than, or equal to, the largest values previously reported. 
In addition, the method can be combined with other heuristic methods, 
and implemented in parallel fashion, allowing it to be applicable to large graphs with more than 10000 nodes. 
\end{abstract} %}}}

%{{{1 pacs
\maketitle 
%}}}1

%{{{1
\section{Introduction} 
\indent Network science has emerged as an important framework to study complex systems~\cite{newman_structure_2006,
caldarelli_scale-free_2007}. One of the most important properties of networks
is the existence of modules/communities; communities are subgraphs of densely inter-connected nodes, and 
nodes in a community are considered to share common 
characteristics~\cite{newman_commpnas_2002,fortunato_community_2010}. Proper community detection allows one to determine potentially hidden relationships between nodes, 
and also allows one to reduce a large complex network into smaller and comprehensible ones. 
For this reason, good community detection within networks has been a subject of great interest. 
There exist various definitions of community~\cite{wasserman1994social,schaeffer2007graph,fortunato_community_2010,ahn_lc_2010}. 
The most widely used approach to detect such sub-groups of nodes with non-random connections involves the use of modularity to
quantify the quality of a given partition of a network~\cite{newman_finding_2004,newmanFastAlgorithm,fortunato_community_2010}. 
Using modularity, the community detection problem is thus recast as a global optimization problem. 
However, finding the \emph{maximum} modularity solution is an NP-hard problem~\cite{brandes2007modularity}, 
and enumeration of all possible partitions is intractable in general. 
Therefore, an efficient optimization algorithm is required to obtain high modularity solutions.
\newline
\indent Most of the modularity optimization studies have focused on developing fast
heuristic methods generating reasonable quality community structures.
Currently, simulated annealing (SA) is considered to be the best algorithm~\cite{fortunato_community_2010,good2010performance} and has been 
adopted in many theoretical and practical studies where communities with high modularity are required~\cite{guimera_functional_2005,sohn2011topological,wang2007search}.
\newline
\indent In this paper, we propose a new modularity maximization method
based on conformational space annealing (CSA) algorithm~\cite{lee1997new,lee1999energy,lee2003unbiased,joo2008multiple,PROT:PROT23059}. 
We show that CSA outperforms SA both in generating better community structures and in computational efficiency.
CSA consistently finds community structures with higher modularity 
using less computational resources. Moreover, for networks containing approximately up to 1000 nodes, 
CSA repeatedly finds converged solutions. Considering the stochastic nature of the algorithm, 
this suggests that the converged solution is likely to be the optimal solution of the network.
%}}}1

%{{{1
\section{Methods}

\begin{figure}
  \includegraphics[width=\columnwidth]{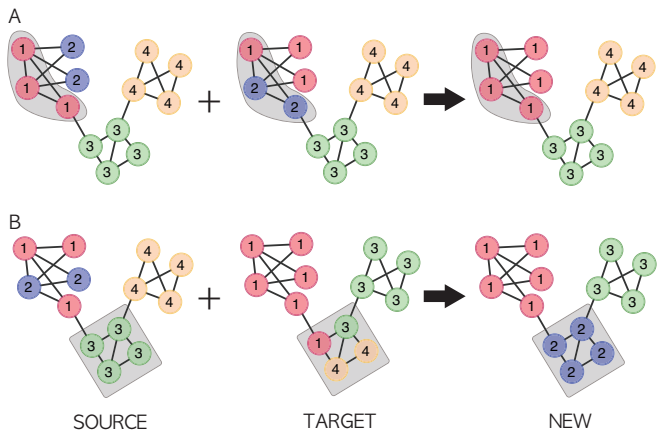}
  \caption{(Color online) Two cross-over operators, (A) convergent copy and (B) divisive copy are shown. 
    In (A) the community indexed by 1 from the source is copied into the target,
    and the new indices are set to 1 or 3 with the probability of $2/3$ and $1/3$. 
    In (B) the community indexed by 3 from the source is copied into the target 
    and the new index 2 is assigned.}
  \label{fig:fig1}
\end{figure}

Let us consider a network with $N$ nodes and $M$ edges.
Modularity measures the fraction of intra-community edges minus its expected value from the null model, 
a randomly rewired network with the same degree assignments. Modularity is defined as 
\begin{equation} 
Q = \sum_{i=1}^{N_c} \left (\frac{l_i}{M}-\left (\frac{D_i}{2M}\right )^2 \right),
\end{equation}
where $N_c$ is the number of assigned communities, $l_i$ is the number of edges within the community $i$ and $D_i$ 
is the sum of degrees of nodes in the community $i$. 
\newline
\indent To benchmark the performance of CSA against that of SA, we implemented SA following existing studies~\cite{guimera_functional_2005,Guimera2005Cartography}.
Initially, using $E=-Q$, a simulation starts at a high temperature $T$, to sample \emph{broad} range of the solution space as well as to avoid trapping in local-minima.
As the simulation proceeds, $T$ is slowly decreased to more completely explore basins of high modularity.
At a given $T$, a set of stochastic movements including $N^2$ single-node moves and $N$ collective moves consisting
 of random merges and splits of communities, are carried out. To split a community, a 'nested' SA method is used~\cite{guimera_functional_2005,Guimera2005Cartography},
 which isolates a target community from the entire network and split it into two communities. 
Each 'nested' SA starts with two randomly separated groups for short annealing and the annealed solution serves as a collective move.
For each trial movement, if $Q$ increases, the movement is accepted, 
otherwise it is accepted with probability $P=exp \left( \frac{{Q_f}-{Q_i}}{T} \right )$.
After a set of movements are tried, $T$ is decreased to $\alpha T$, where $\alpha=0.995$.
\newline
\indent Our method, CSA, is a global optimization method which combines essential ingredients of three methods: Monte Carlo with minimization (MCM)~\cite{Li1987monte},
 genetic algorithm (GA)~\cite{goldberg1989genetic}, and SA~\cite{Kirkpatrick1983optimization}. As in MCM, 
we consider only the phase/conformational space of local minima; i.e., 
all solutions are minimized by a local minimizer. As in GA, we consider many solutions 
(called \emph{bank} in CSA) collectively, and we perturb a subset of bank solutions by cross-over between solutions and mutation. Finally, as in SA, we introduce a parameter $D_{cut}$, 
which plays the role of the temperature in SA. In CSA, each solution is assumed to represent 
a hyper-sphere of radius $D$ in the solution space. Diversity of sampling is directly controlled by 
introducing a distance measure between two solutions and comparing it with $D_{cut}$, to deter two solutions from coming 
too close to each other. Similar to the reduction of $T$ in SA, the value of $D_{cut}$ is slowly reduced in CSA; hence the name conformational space annealing.
\newline
\indent To apply CSA to optimize modularity, three ingredients are required:
(a) we need a local modularity maximizer for a given network partition, (b) we need a distance measure between two $Q$-maximized network partitions, 
and (c) we need ways to combine two parent partitions to generate a daughter partition which will be $Q$-maximized subsequently. 
\newline
\indent Here, the community structure is represented by assigning an index to each node, where nodes with an identical 
index belong to the same community. For local maximization of $Q$, we use a quench procedure 
which accepts a move if and only if it improves $Q$, equivalent to SA at $T=0$.
\newline
\indent The distance between two community structures is measured by the variation of information (VI)~\cite{meila2007comparing}.
For two given partitions of $X$ and $Y$, $VI$ is defined as 
\begin{align*}
  VI(X,Y) = H(X,Y)-I(X;Y)
\end{align*}
where $H$ is the entropy function and $I$ is the mutual information function of the probability 
$p(x,y) = n_{x,y}/n$, where $n$ is the number of total nodes, $x$/$y$ refers to a community from $X$/$Y$, and 
$n_{x,y}$ is the number of nodes shared by $x$ and $y$. 
With $H$ and $I$ defined by
\begin{align*}
  H(X,Y) & = -\sum_{x,y}p(x,y) \log p(x,y) \\
         & = -\sum_{x,y} \frac{n_{x,y}}{n} \log \left( \frac{n_{x,y}}{n} \right),
\end{align*}

\begin{align*}
  I(X;Y) & = \sum_{x,y}p(x,y)\log \left(\frac{p(x,y)}{p(x)p(y)}\right) \\
         & = \sum_{x,y}\frac{n_{x,y}}{n}\log\left(\frac{n_{x,y}n}{{n_x}{n_y}}\right),
\end{align*}
$VI$ can be reduced to
\begin{equation}
  VI(X,Y) = -\frac{1}{n}\sum_{x,y}n_{x,y}\log\left(\frac{n^{2}_{x,y}}{n_{x}n_{y}}\right),
\end{equation}
where $p(x)=n_{x}/n$ and $n_{x}$ is the number of nodes in community $x$. 
If $X$ is identical to $Y$, it is easy to show that $VI(X,Y)=0$.
We have also tried other measures such as Rand index~\cite{rand1971objective} 
and normalized mutual information (NMI)~\cite{jain2003robust} and they gave no significant difference in results.
\newline
\indent In CSA, we first generate 50 random partitions which are subsequently maximized by quench procedures.
We call these solutions as the \emph{first bank} which is kept unchanged during the optimization.
We make a copy of the first bank, and call it the \emph{bank}. The partitions in the bank are 
updated by better solutions found during the course of optimization. 
The initial value of $D_{cut}$ is set as $D_{avg}/2$, where $D_{avg}$ is the average
distance between partitions in the first bank. 
A number of partitions (30 in this study) in the bank are selected as \emph{seed} partitions.
With each seed, 20 trial partitions are generated by cross-over between the seed and a randomly chosen partition 
from either the bank or the first bank. An additional 5 are generated by random mutation of the seed.
\newline
\indent For a cross-over, we use two operations, a convergent copy and a divisive copy, shown in Figure~\ref{fig:fig1}. 
In both cases, one community represented by an index is randomly selected from a source solution and it is copied into 
a target solution after assigning a new index. For the convergent copy, the new index is chosen from one of the 
neighboring indices of the copied nodes from the target in a random fashion. For the divisive copy, a new index not present
 in the target is chosen. The rationale of using these operators is that the community index itself has no particular meaning,
 while a well-defined community structure from one solution can serve as an advantageous feature that should be preserved 
to generate a better solution. 
For each operation, the minimum number of nodes that should be copied are randomly determined between 1\% to 40\% of total nodes 
and the above operation is repeated until the total number of copied nodes exceeds the number.
\newline
\indent For mutation, random merge and split operators were introduced. The random merge was carried out by combining two adjacent communities.
The random split operator divides a community into two randomly assigned groups.
All trial conformations generated by cross-over and mutation operations are optimized by quench procedures.
It should be noted that only \emph{local} moves are used in the quench procedures 
since the divergent and divisive copy operators can act as the merge and split moves used in SA.

\indent After local-maximization of trial partitions, these partitions are used to update the bank. 
The modularity of a trial partition $\alpha$ is compared with the modularities of partitions in the bank. 
If $\alpha$ is worse than the worst partition of the bank, it is discarded. 
Otherwise, we find the partition $A$ in the bank which is closest to $\alpha$, as determined by distance $D(\alpha,A)$.
If $D(\alpha,A)<D_{cut}$, $\alpha$ is considered as similar to $A$ and it replaces $A$ if $\alpha > A$. If $\alpha < A$ it is discarded. 
If $D(\alpha,A)>D_{cut}$, $\alpha$ is regarded as a new partition similar to none in the bank, 
and it replaces the worst existing partition, that is, it replaces the lowest modularity partition in the bank. 
We carry out this operation for all trial partitions. 
With updated bank, new seed partitions are selected again from the bank which have not yet been used as seeds.
This entire process of generating partitions by perturbation and subsequent local maximization and updating bank 
is repeated until all partitions in the bank are used as seeds. 
At each iteration step, $D_{cut}$ is reduced with a pre-determined ratio. 
After $D_{cut}$ reaches to its final value, $D_{avg}/5$, it is kept constant. 
\newline
\indent Once all partitions in the bank are used as seeds without generating better partitions, 
implying that the procedure might have reached a deadlock, we reset all bank partitions to be eligible for seeds and 
repeat another round of search procedure. After this additional search also reaches a deadlock, we expand our search space by adding an additional 50 randomly generated
and optimized partitions to the bank, and repeat the whole procedure. 
In this study, we terminated our calculation after 100 partitions were used as seeds. 
Additional adding cycles should be considered for more rigorous optimization, especially for problems with high complexity.
%}}}1

%{{{1
\section{Results}
\begin{center}
  \begin{table}
    {\small
      \begin{tabular}{ccc}
        \hline
        Network & Nodes & Edges \\
        \hline
        Dolphins           & 62    & 159   \\
        Les Mis\`erables   & 77    & 254   \\
        Political books    & 105   & 441   \\
        College football   & 115   & 613   \\
        Jazz               & 198   & 2742  \\
        USAir97            & 332   & 2126  \\
        Netscience\_main   & 379   & 914   \\
        C. \emph{elegans}  & 453   & 2025  \\
        Electronic Circuit (s838) & 512 & 819 \\
        E-mail             & 1133  & 5451  \\
        Erdos02            & 6927  & 11850 \\
        PGP                & 10680 & 24316 \\
        condmat2003        & 27519 & 116181\\
        \hline
      \end{tabular}
    }
    \hfill
    \caption{Number of nodes and edges of benchmark networks used in this study are displayed.}
    \label{tb:t0}
  \end{table}
\end{center}

\indent To compare performance of CSA and SA, we applied CSA and SA to a number of real-world networks commonly used in existing modularity optimization studies, shown in Table~\ref{tb:t0}.
All networks considered are undirected and unweighted.
Due to the stochastic nature of both methods, we performed 50 independent simulations for each method. 
The results are summarized in Table~\ref{tb:t1}.
The maximum, average and standard deviation of modularity values obtained by both methods are displayed. 
As a measure of required computational resources, we counted the number of function evaluations performed until the maximum modularity solution is found, $N^{max}$.
We observe that CSA consistently finds equal or higher modularity solutions than does SA 
for all networks tested, with a smaller number of function evaluations.
To demonstrate the search efficiency of CSA more clearly, we also measured the number of function evaluations required by CSA 
to generate a solution equivalent to the best modularity obtained by SA, which is denoted as $N^{equal}_{CSA}$ in Table~\ref{tb:t1}.
CSA clearly requires many fewer function evaluations to generate solutions better than the best ones obtained by SA.
For small networks (e.g. up to the Jazz musician network), CSA finds the best solution with less than 10\% of the function evaluations required by SA 
and for the worst case, the C. \emph{elegans} network, CSA requires only 25\% of the function evaluations of SA.
\begin{center}
  \begin{table*}
    {\small
      \begin{tabular}{ccccccccccccc}
        \hline
        & & \multicolumn{3}{c}{CSA} & & \multicolumn{3}{c}{SA} \\ \cline{3-5} \cline{7-9}
        Network & & $Q_{max}$ & $Q_{avg}$ & $\sigma$ & & $Q_{max}$ & $Q_{avg}$ & $\sigma$ & $N^{max}_{CSA}/N^{max}_{SA}$ & $N^{equal}_{CSA}/N^{max}_{SA}$ & $\emph{t}_{CSA}$ & $\emph{t}_{SA}$\\
        \hline
        Dolphins                      & & 0.52852 & 0.52852 & 0        & & 0.52852 & 0.52507 & 0.0036 & 0.077  & 0.077  & 0.09   & 0.74    \\
        Les Mis\`erables              & & 0.56001 & 0.56001 & 0        & & 0.56001 & 0.55194 & 0.0071 & 0.362  & 0.362  & 0.01   & 0.18    \\
        Political books               & & 0.52724 & 0.52724 & 0        & & 0.52724 & 0.52723 & 0      & 0.055  & 0.019  & 0.07   & 2.52    \\
        College football              & & 0.60457 & 0.60457 & 0        & & 0.60457 & 0.60457 & 0      & 0.093  & 0.093  & 0.05   & 0.26    \\
        Jazz                          & & 0.44514 & 0.44514 & 0        & & 0.44487 & 0.44477 & 1.6e-4 & 0.073  & 0.052  & 0.17   & 679.4   \\
        USAir97                       & & 0.36824 & 0.36824 & 0        & & 0.35376 & 0.34787 & 0.0044 & 0.271  & 0.010  & 0.13   & 429.2   \\
        Netscience\_main              & & 0.84859 & 0.84859 & 0        & & 0.84383 & 0.83544 & 0.0044 & 0.345  & 0.019  & 1.3    & 263.3   \\
        C. \emph{elegans}             & & 0.45325 & 0.45325 & 0        & & 0.45212 & 0.44927 & 0.0026 & 0.960  & 0.246  & 16.8   & 2512.3  \\
        Electronic Circuit (s838)     & & 0.81936 & 0.81936 & 0        & & 0.81871 & 0.80812 & 0.0048 & 0.639  & 0.424  & 2.6    & 1129.4  \\
        E-mail                        & & 0.58283 & 0.58282 & 2.2e-5   & & 0.58198 & 0.58015 & 0.0015 & 0.510  & 0.119  & 73.6   & 42296   \\
        Erdos02                       & & 0.71843 & 0.71782 & 3.2e-4   & &   -     &   -     &   -    &  -     &  -     & 3356   &   -     \\
        PGP                           & & 0.88675 & 0.88648 & 1.1e-4   & &   -     &   -     &   -    &  -     &  -     & 10757  &   -     \\
        condmat2003                   & & 0.76745 & 0.76484 & 0.0010   & &   -     &   -     &   -    &  -     &  -     & 57609  &   -     \\
        \hline
    \end{tabular}}
    \hfill
    \caption{Modularity optimization results obtained by 50 seperate runs of Mod-CSA and SA are displayed. $Q_{max}$ denotes the maximum modularity value, $Q_{avg}$ the average of maximum modularity value of each run, $\sigma$ the standard deviation of the modularity value, $N^{max}$ the number of function evaluations until the calculation reached the maximum modularity, $N^{equal}_{CSA}$ the number of function evaluations required for CSA runs to sample equal or higher modularity solutions than the maximum modularity of SA runs, $\emph{t}$ the average execution time to find the best solution in seconds. All simulations were performed on Intel Nehalem core @ 2.93GHz. CSA and SA runs were performed with 64 cpus and single cpu, respectively. For the three largest networks, SA results are not available because calculations were not finished within 2 days.}
    \label{tb:t1}
  \end{table*}
\end{center}
\begin{center}
  \begin{table*}
    {\small
      \begin{tabular}{ccccccc}
        \hline
        & \multicolumn{2}{c}{CSA} & &\\ \cline{2-3}
        Network & $N_{c}$ & $Q_{max}$ & $Q_{pub}$ & $Q_{opt}$ & $\%_{opt}^{SA}$ & Source \\
        \hline
        Dolphins                  & 5   & 0.52852 & 0.5285 & 0.5285 & 16.0  & ~\cite{xu2007finding,agarwal2008modularity,aloise2010column}   \\
        Les Mis\`erables          & 6   & 0.56001 & 0.5600 & 0.5600 & 20.0  & ~\cite{aloise2010column}   \\
        Political books           & 5   & 0.52724 & 0.5272 & 0.5272 & 100.0 & ~\cite{agarwal2008modularity,Noack2009multi,aloise2010column}  \\
        College football          & 10  & 0.60457 & 0.6046 & 0.6046 & 100.0 & ~\cite{agarwal2008modularity,ye2008adaptive,aloise2010column}  \\
        Jazz                      & 4   & 0.44514 & 0.4451 &   -    &  -    & ~\cite{schuetz2008efficient,Noack2009multi,agarwal2008modularity,duch2005community}  \\
        USAir97                   & 6   & 0.36824 & 0.3682 & 0.3682 & 0.0   & ~\cite{aloise2010column}   \\
        Netscience\_main          & 19  & 0.84859 & 0.8486 & 0.8486 & 0.0   & ~\cite{aloise2010column}   \\
        C. \emph{elegans}         & 9   & 0.45325 & 0.452  &   -    &  -    & ~\cite{liu2010advanced}    \\
        Electronic Circuit (s838) & 16  & 0.81936 & 0.8194 & 0.8194 & 0.0   & ~\cite{aloise2010column}   \\
        E-mail                    & 10  & 0.58283 & 0.582  &   -    &  -    & ~\cite{liu2010advanced}    \\
        Erdos02                   & 40  & 0.71843 & 0.7162 &   -    &  -    & ~\cite{Noack2009multi}     \\
        PGP                       & 100 & 0.88674 & 0.8841 &   -    &  -    & ~\cite{Noack2009multi,liu2010advanced}     \\
        condmat2003               & 80  & 0.76745 & 0.761  &   -    &  -    & ~\cite{ye2008adaptive}     \\
        \hline
    \end{tabular}}
    \hfill
    \caption{Comparison between the maximum modularity values obtained by CSA, $Q_{max}$,
      with previously published ones, $Q_{pub}$, and the maximum values obtained by 
      the exact method~\cite{aloise2010column}, $Q_{opt}$, is displayed. $N_c$ denotes the number of communities found by CSA. 
      Source indicates the reference that the modularity value is collected. $\%_{opt}^{SA}$ denotes 
      the percentage of SA runs that reached to the optimal modularity community structure.}
    \label{tb:t2}
  \end{table*}
\end{center}

\begin{figure}
  \includegraphics[width=\columnwidth]{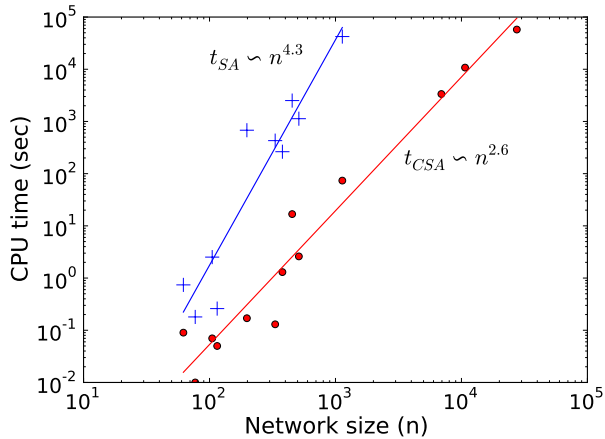}
  \caption{(Color online) Comparison of time complexities of CSA (\emph{red, circle}) and SA (\emph{blue, cross}) is shown.
    Network size, n, represents the number of nodes,
    and CPU time corresponds to the average time to find final solutions in seconds.}
  \label{fig:fig2}
\end{figure}

\indent It should be noted that CSA can be applied to networks containing more than $10^{3}$ nodes where for SA this is impractical.
For the three largest networks in Table~\ref{tb:t1}, CSA found good solutions within a reasonable computational time 
whereas SA runs did not yield reasonable value of modularity within 2 days of wall clock time.
This difference in computational time reflects a number of factors. It is partly due to the high parallel efficiency of the CSA algorithm~\cite{lee2000efficient}.
In SA, generation of a trial solution is dependent on its previous state, which makes it impractical to implement the algorithm in a parallel fashion.
However, the majority of computation in CSA consists of independent local maximization procedures on hundreds of trial solutions 
generated by cross-over and mutation, and each maximization can be separately carried out.
The quench procedure in CSA consists of local moves only, which is rather fast with large networks. 
On the other hand, the most time-consuming operation in SA is the splitting move by the nested SA procedure 
which we find is indeed essential to obtain good SA solutions. 
In CSA, the operation of the divisive copy when generating trial solutions plays the equivalent role of the split move in SA.
To compare the computing efficiency of CSA with existing methods, the time complexities of CSA and SA are estimated 
based on the simulation results with the benchmark networks. As shown in Figure~\ref{fig:fig2},
the time complexity of CSA is estimated to be $O(n^{2.6})$ which is comparable to
other heuristic methods~\cite{danon_comparing_2005,lancichinetti2009community} and better than that of SA, $O(n^{4.3})$, where $n$ is the number of nodes.

\indent In terms of convergence, CSA yields more robust solutions than SA.
Except the political books and college football networks, the maximum modularity solution found by SA varied from simulation to simulation. 
For networks containing over 300 nodes, SA failed to sample the optimal solution, 
which raises serious concerns when applying SA to modularity optimization for practical use~\cite{good2010performance}.
However, for all test networks up to about $10^3$ nodes, all CSA runs converged to the same solution, except the E-mail network where 41 out of 50 converged.
Considering the small size of networks and the stochastic nature of the algorithm, 
we believe that the converged solution of each network is likely to be the true maximum modularity of the network. 
\newline
\indent We also compared maximum modularities obtained by CSA with the maximum values from previous publications; see Table~\ref{tb:t2}.
CSA finds equivalent or higher $Q$ values compared to existing studies in all networks tested.
\newline
\indent Recently, the exact maximum modularity values of several small benchmark networks up to 512 nodes were reported; they 
are displayed in Table~\ref{tb:t2} as $Q_{opt}$~\cite{aloise2010column}. 
We performed 50 independent runs for these networks and all runs converged to the optimal solutions \emph{without exception}. 
This result supports the hypothesis that CSA is efficient enough to find the putative maximum modularity solution for a network containing up to $10^3$ nodes.
\newline
\indent CSA algorithm presented in this work aims to obtain optimal modularity solutions and the method is not 
free from the problem of the resolution limit arising from using modularity~\cite{fortunato_resolution_2007}.
However, the CSA procedure and operators proposed in this work are general, and 
can be used to optimize other fitness functions.
To overcome the resolution limit issue, more robust fitness functions should be considered to be combined with CSA,
such as the map equation~\cite{Rosvall29012008} or the partition density~\cite{ahn_lc_2010}. 
It should be noted that the current work can be extended to deal with directed or weighted networks 
in conjunction with modified modularity functions~\cite{newman2004analysis,arenas_generalmod_2007}.
In order to handle large networks, CSA can be combined with other efficient heuristics, such as the fast unfolding method~\cite{blondel2008fast}, 
instead of the stochastic quench procedure used in this study.
\newline
%%}}}1

%%{{{1
\section{Conclusion}
In this paper, we propose a new modularity optimization method based on conformational space annealing algorithm, Mod-CSA.
Compared to SA, our method is faster. Further, while it finds equivalent modularity partitions for relatively small networks,
 for the larger more challenging ones, it typically finds higher modularity partitions.
For small networks consisting up to $10^3$ nodes, despite its stochastic nature, Mod-CSA solutions converge to an identical solution, 
which appears to be the best solution possible; this is not
possible in other stochastic algorithms. Mod-CSA can be implemented in a highly parallel fashion and is thus applicable to large networks
where SA is not. In addition, Mod-CSA can be extended to deal with large networks by using fast heuristic methods 
as a local optimizer. 
%}}}
\begin{acknowledgments}
The authors acknowledge support from Creative Research Initiatives (Center for In Silico Protein Science, 20110000040) of MEST/KOSEF. 
We thank Korea Institute for Advanced Study for providing computing resources (KIAS Center for Advanced Computation Linux Cluster) for this work.
\end{acknowledgments}

%%\bibliography{all}

%%\input{paper_main.bbl}

%

\end{document}